\documentclass[runningheads]{llncs}
\usepackage[T1]{fontenc}
\usepackage{lmodern}
\usepackage{graphicx,booktabs,tabularx,array,amsmath,amssymb,xcolor,url,tikz}
\usetikzlibrary{arrows.meta,positioning}
\usepackage[hidelinks]{hyperref}
\begin{document}
\title{Where Cyber Agents Struggle: Bottleneck Analysis of Multi-Stage LLM Agents}
\titlerunning{Where Cyber Agents Struggle}
\author{Saeedeh Lohrasbi\inst{1} \and Mohammad Mamun\inst{1} \and
Ahmed Yehia\inst{1} \and Scott Buffett\inst{1} \and Sherif Saad\inst{2}}
\authorrunning{S. Lohrasbi et al.}
\institute{Cybersecurity, DTRC, National Research Council, Ottawa, Canada\\
\email{\{saeedeh.lohrasbi,mohammad.mamun,}\\
\email{ahmed.abdalmagid,scott.buffett\}@nrc-cnrc.gc.ca}
\and University of Windsor, Windsor, Ontario, Canada\\
\email{sherif.saad@uwindsor.ca}}
\maketitle

\begin{abstract}
Multi-stage LLM-based cyber agents may complete attack workflows while remaining brittle, costly, or reliant on incorrect interpretations of execution evidence. Success rates alone obscure inefficiency, adaptation through retries, and recognition of success or failure. We present an end-to-end diagnostic study of an Autonomous Adversary system with orchestrator, executor, and validator LLMs in enterprise-like lateral-movement scenarios. Six frontier models are evaluated across two scenarios and three modes: expert-defined, self-scaffolded, and fully autonomous. We assess validator consistency and evidence grounding; introduce a subtask-conditioned, cost-aware score for abnormal token use, retries, and runtime; and use comparative LLM-as-a-Judge analysis to identify planning deficiencies, including tool misalignment, plan similarity, over-specification, inadequate probing, and weak recovery. Validators are generally relevant and evidence-grounded but often nonspecific and overly optimistic. Bottlenecks cluster in credential and lateral-movement tasks, spread with scenario complexity, and vary more under full autonomy. Reliable evaluation must assess outcomes, evidence interpretation, resource use, and adaptation after failure.

\end{abstract}

\section{Introduction}
Large language models (LLMs) support agents that plan, invoke tools, interpret observations, and revise actions~\cite{yao2022react}. Cybersecurity workflows link credential acquisition, identity transition, lateral movement, and cleanup. Command execution need not establish the state required by the next stage; eventual success may conceal costly repetition. A final success indicator captures neither distinction.

Interactive benchmarks, including InterCode and CyBench, provide useful environments for evaluating execution-grounded capabilities~\cite{yang2023intercode,zhang2025cybench}. Our question is complementary: within an observed cyber-agent execution, where does progress become constrained, and what persistent weakness best explains that constraint? Answering it requires separating resource consumption from behavioral interpretation, and the acting model from the orchestration and validation components surrounding it.

The comparison unit depends on task structure. Shared decompositions support aligned subtask comparisons; agent-generated objectives do not. We combine \emph{task-level localization} of relative resource amplification under expert decomposition with \emph{trajectory-level diagnosis} of persistent progress-limiting behavior across all scaffolding modes.
We apply this framework to selected high-performing trajectories from two lateral-movement scenarios, six acting models, and three modes, with incomplete configuration coverage. Five judges assess complete traces, while a separate assessment examines validator explanations against execution evidence.

Our contributions are: (1) a two-resolution framework connecting resource localization with trajectory diagnosis; (2) separation of primary bottlenecks, mechanisms, recovery, and framework contributions; and (3) a case study revealing scenario-dependent relations between scaffolding and diagnosed bottlenecks. The contribution is diagnostic, not a representative model ranking or an experimental demonstration of causality.
The data and source code are available at \url{doi.org/10.4224/40004009} and \url{https://github.com/wherecyberagentsstruggle/LLM_as_Judge} respectively. 

\section{Related Work}
\paragraph{Cyber agents and interactive evaluation.}
AutoAttacker studies LLM-guided post-breach operations in simulated networks~\cite{xu2024autoattacker}; CurriculumPT explores curriculum-guided task scheduling for autonomous penetration testing~\cite{wu2025curriculumpt}. InterCode standardizes interaction with execution feedback~\cite{yang2023intercode}, while CyBench evaluates cybersecurity capabilities through capture-the-flag tasks~\cite{zhang2025cybench}. These efforts motivate evaluating multi-step tool use, but a capability score and a diagnosis of persistent difficulty answer different questions. Our analysis concentrates on the latter, within two enterprise-like lateral-movement workflows rather than a broad benchmark suite.

\paragraph{Failure taxonomies and diagnostic judgment.}
Winston and Just classify failures in tool-augmented LLM systems~\cite{winston2025taxonomy}. Such taxonomies help separate faults in reasoning, tool interaction, and the surrounding system. We combine categorical diagnosis with resource signals and distinguish an isolated error from a weakness that persists across dependent actions. The task-level and trajectory-level labels serve different purposes: local plan properties need not coincide with the dominant explanation for the full run.

LLM-as-a-Judge enables scalable assessment but introduces evaluator bias and disagreement~\cite{zheng2023judging}. We use multiple judges to expose disagreement, preserving ties and interpreting consensus as consistency rather than human-validated correctness.

\section{Experimental System and Study Design}
\subsection{Execution Architecture and Scaffolding}
The Autonomous Adversary operates in a controlled range with segmented subnets, remote-administration services, a root domain and subdomains, domain controllers, domain-joined workstations, and security misconfigurations. There is no active defender or execution-blocking detection penalty. The resulting observations concern planning, execution, state interpretation, and recovery in this setting, not resilience against adaptive defensive intervention.

\begin{figure}[t]
\centering
\includegraphics[width=\linewidth]{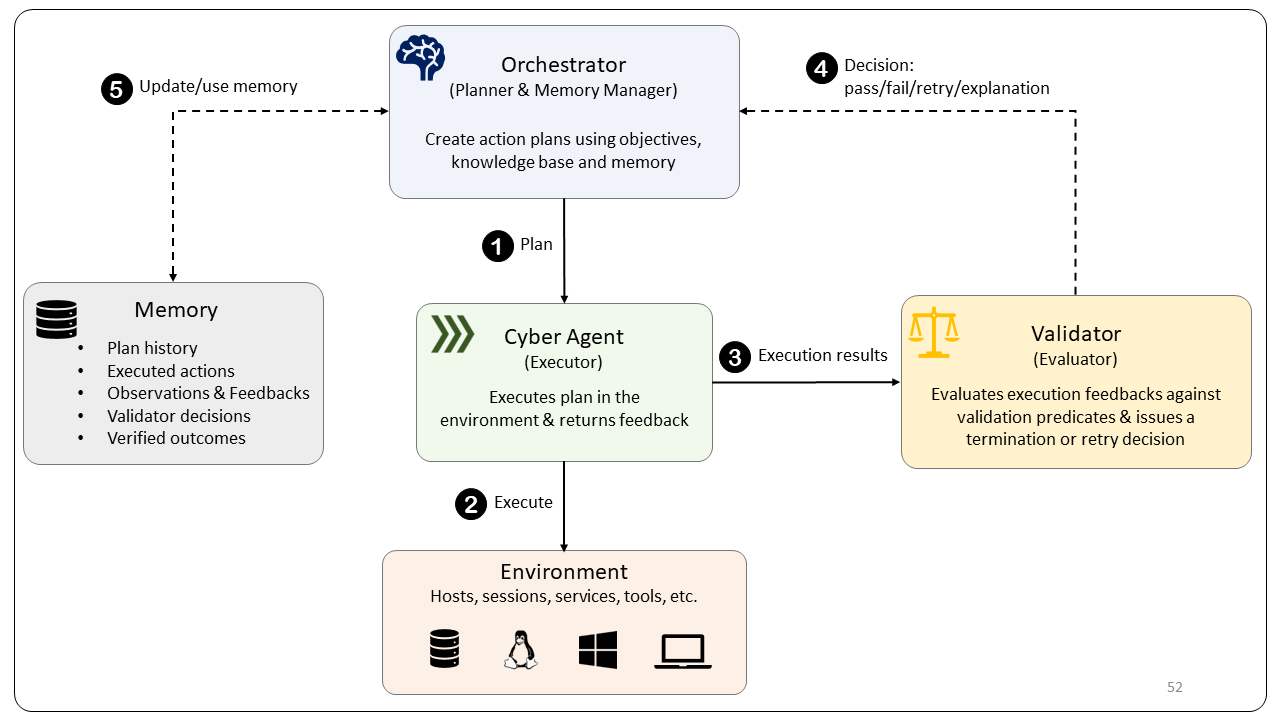}
\caption{Autonomous Adversary execution architecture and evidence flow used for bottleneck analysis.}
\label{fig:architecture}
\end{figure}

Figure~\ref{fig:architecture} separates the execution loop from its offline analysis. The acting LLM serves as the \emph{orchestrator}, generating operational plans and maintaining execution context. The \emph{cyber executor} runs proposed commands and tool invocations, returning standard output, standard error, exit status, tool responses, and observable side effects. The \emph{system validator} compares this evidence with the applicable success criterion and determines whether execution continues, retries, submits, or halts. Memory includes plan history, actions, observations, validator decisions, and recorded outcomes; a validator-accepted outcome is not necessarily independently verified.

The system is based on the Autonomous Adversary framework developed by Mamun et al.~\cite{mamun2026autonomous}. The framework maintains memory: after each execution, the LLM extracts information learned from the task and adds it to its knowledge base.
The reported traces involve shell and PowerShell commands, Windows Management Instrumentation Command-line (WMIC), Mimikatz, and Caldera-agent deployment. This identifies observed execution mechanisms, not the complete interface exposed to the acting models.
There is no fixed tool list. Agents generate abilities that the cyber executor runs through PowerShell on Windows or \texttt{sh} on Linux/macOS, using utilities available on the target. All evaluated models use the same structured action interface; the three modes differ only in planning and context.

The three modes differ in who supplies the decomposition. In \emph{expert-defined} mode (E), a human evaluator supplies ordered subtasks and verification criteria. In \emph{self-scaffolded} mode (S), the acting model generates subtasks and intermediate success criteria from a high-level objective. In \emph{fully autonomous} mode (A), it receives the objective and information observable from its initial foothold, performs reconnaissance, maintains a security-state representation, and determines intermediate objectives and execution order.
The expert-defined subtasks and verification criteria were developed by an offensive cybersecurity expert specializing in red teaming and penetration testing, who independently planned the scenarios, solved them manually, and decomposed them into tasks.

\subsection{Scenarios, Budgets, and Selection}
Scenario~1 (S1) begins at a low-privilege workstation with an active administrator session. Its nine expert-defined subtasks cover execution-context setup, agent deployment, credential acquisition, identity transition, protected-resource access, and restoration of the baseline state. Scenario~2 (S2) begins with an administrative foothold on a jump server. Its ten expert-defined subtasks involve an intermediate workstation, credential recovery and reuse, a writable domain-controller resource, domain-administrator execution, restricted-file retrieval, and artifact removal. These scenarios share a range but differ in their dependency structure and starting privileges.

The evaluated acting models include Claude Opus~4.5, Claude Sonnet~4.5,
Gemini~3 Pro Preview, GPT-5.1, GPT-4o-mini, and DeepSeek-V3.2-Speciale.
For cross-setting comparison, we focus on Claude Opus~4.5, Claude
Sonnet~4.5, Gemini~3 Pro Preview, and GPT-5.1, which provide the most
consistent trajectory coverage across the six scenario--mode settings.
Table~\ref{tab:coverage} reports this four-model comparison cohort.
The Claude Opus~4.5 S2-A judge panel is incomplete (judge call was blocked by a provider-level cybersecurity safety filter) and is therefore
excluded from the corresponding aggregate comparisons.

\begin{table}[t]
\caption{Primary trajectory diagnoses for the four-model comparison cohort.
PL: planning and strategy; AE: action execution; RA: recovery and adaptation;
VF: validation or framework. Fractions are modal votes out of five, not success
rates. Slashes denote ties. INC denotes an incomplete judge panel excluded
from aggregates.}
\label{tab:coverage}
\centering\scriptsize
\setlength{\tabcolsep}{2.3pt}
\begin{tabular}{lcccccc}
\toprule
Acting model & S1-E & S1-S & S1-A & S2-E & S2-S & S2-A\\
\midrule
Opus 4.5
& AE 4/5 & PL 5/5 & RA 4/5 & PL 4/5 & AE/PL 2/5 & INC\\

Sonnet 4.5
& AE/PL 2/5 & PL 4/5 & PL 3/5 & PL 3/5 & AE 3/5 & PL 3/5\\

Gemini 3 Pro Preview
& AE 2/5 & PL 4/5 & PL 5/5 & RA 4/5 & AE 4/5 & PL 3/5\\

GPT-5.1
& AE 5/5 & PL 5/5 & PL/VF 2/5 & PL/RA 2/5 & AE 4/5 & PL 3/5\\
\bottomrule
\end{tabular}
\end{table}

A per-call reference of 45,000 tokens was chosen to support long contexts and standardize token-efficiency comparisons. This value covers the input and generated output of one LLM call, not the entire run. Reaching the output limit stops generation and may truncate the response. Such truncation can constrain an individual action or recovery step; it does not establish that the entire run has exhausted a token budget. We do not isolate a budget effect experimentally.

For each available model--scenario--mode configuration, the run with the greatest judge-verified task progress was selected for analysis. This selection targets bottlenecks that persist in the strongest observed executions and represents best-case observed performance rather than average performance.
DeepSeek was omitted from some settings because of upstream provider errors. GPT-4o-mini was omitted from some settings because it failed to produce the structured output required by the cyber executor and made little progress.
Selection favors stronger observed executions and cannot establish bottleneck prevalence across repeated runs. The repeated validator-calibration runs below are a separate sample, not a robustness study of these diagnoses.

\section{Multi-Resolution Diagnostic Method}
\subsection{Task-Level Localization}
A \emph{run} executes a scenario; a \emph{subtask} is an intermediate stage; a \emph{plan} proposes commands or tools; an \emph{attempt} executes a plan. Retrying repeats or repairs a plan, while replanning changes strategy. For aligned expert-defined subtask $s_i$, $N_{m,i}=\sum_k n_{m,i,k}$ counts attempts across plans for model $m$. We measure tokens, elapsed time, and $N_{m,i}$.

For resource signal $\phi$, normalize across models with valid observations for the \emph{same} subtask, using mean $\mu_i^{(\phi)}$ and standard deviation $\sigma_i^{(\phi)}$:
\begin{equation}
 z_{m,i}^{(\phi)}=\frac{x_{m,i}^{(\phi)}-\mu_i^{(\phi)}}{\sigma_i^{(\phi)}},\qquad
 B_m(s_i)=\alpha[z_{m,i}^{(\mathrm{tokens})}]_+
 +\beta[z_{m,i}^{(N)}]_++\gamma[z_{m,i}^{(\mathrm{time})}]_+,
\label{eq:score}
\end{equation}
where $[z]_+=\max(0,z)$ and the weights are nonnegative. The formula applies to signals with nonzero cross-model variance. A high score indicates relative resource amplification, including in eventually successful subtasks. A zero score does not establish success or intrinsic ease; an expensive subtask shared equally across models may have no positive standardized deviation.

We set $\alpha=\beta=\gamma=1$, giving equal weight to the three positive standardized resource signals.
The score is cohort-dependent and is not an absolute difficulty scale. Its value can change if the comparison models or weights change. We report it only within the shared expert-defined decomposition, not across agent-generated subtask indices. No empirical weight or normalization sensitivity study is claimed.

GPT-5 Codex evaluates observable candidate plans, returning category indicators, a primary label, severity, and an evidence-based explanation. The seven local labels are: \emph{plan similarity} (superficial strategic variation); \emph{single-point dependency} (a shared failure-critical assumption); \emph{over-specification} (unverified environmental details); \emph{missing environmental probing} (omitted prerequisite checks); \emph{lack of recovery paths}; \emph{tool misalignment}; and \emph{inefficient planning depth} (redundant steps). These describe visible plan structure, not hidden model reasoning. A single planning judge supplies an interpretation rather than a validated causal attribution.

\subsection{Trajectory Taxonomy and Evidence}
Across all modes, the full trace contains the initial objective and context, followed by plans, actions, execution evidence, and validator decisions. It preserves retries, revisions, state updates, host and session transitions, and outcomes.

A trajectory bottleneck persistently constrains progress or causes avoidable resource use. An isolated command failure qualifies only if its cause persists or blocks a dependency. Judges distinguish the visible \emph{symptom}, the \emph{immediate cause}, and the hypothesized \emph{root cause} explaining persistence. Root causes are diagnostic interpretations, not intervention-established effects.

\begin{table}[t]
\caption{Primary trajectory-level categories. One category is selected by each judge; mechanisms and contributing causes are recorded separately.}
\label{tab:taxonomy}
\centering\small
\begin{tabularx}{\linewidth}{@{}p{31mm}X@{}}
\toprule
Category & Operational distinction\\
\midrule
Planning and strategy & Incomplete, incorrectly ordered, or assumption-dependent strategy; omitted reconnaissance or prerequisites.\\
Action execution & Viable strategy incorrectly realized through commands, tools, host, identity, session, or syntax.\\
State interpretation and tracking & Incorrect belief or update about successes, failures, acquired resources, or unknowns.\\
Recovery and adaptation & Useful failure evidence is available, but the operative hypothesis or strategy is not revised.\\
Validation or framework & Decisions, criteria, observability, orchestration, or interfaces distort or block progress.\\
Environment or infrastructure & External connectivity, services, system state, or infrastructure principally causes blockage.\\
Other & None of the above adequately captures the deepest progress-limiting cause.\\
\bottomrule
\end{tabularx}
\end{table}

Each judge selects one primary category (Table~\ref{tab:taxonomy}), at most two contributing bottlenecks, and specific mechanism tags. Local planning labels can support, but are not interchangeable with, trajectory categories.

\emph{Meaningful adaptation} changes the operative hypothesis, resolves a prerequisite, gathers discriminating evidence, changes dependencies, or adopts a substantively different strategy. Changing syntax, wrappers, tools, protocols, or hostnames while preserving a failed assumption is repetition, not meaningful recovery. A separate field records material validator, framework, or environmental contribution. Such a contribution can coexist with an agent-side primary bottleneck; it is not an additional mutually exclusive category.

\subsection{Evaluator Roles and Aggregation}
Table~\ref{tab:evaluators} distinguishes the four evaluation components. The trajectory panel consists of DeepSeek V4 Pro, GPT-5.6 Sol, Kimi K3, Gemini 3.6 Flash, and GPT-5.6 Luna Pro, queried separately through OpenRouter.
Judge selection considered coding and reasoning capabilities, model diversity, context capacity, availability, and preliminary trials. GPT-5 Codex was the initial choice for plan assessment because the cyber tasks involve coding, command construction, and tool invocation, motivating a judge capable of identifying implementation and execution errors. 

Within each configuration, the system validator used the same model as the acting agent. Comparisons therefore concern the acting-agent/validator combination, while the two roles remain distinct.
Using several judges exposes disagreement between evaluators, but separate calls do not imply statistically independent or unbiased judgments.

\begin{table}[t]
\caption{Evaluation components, inputs, and roles. The online validator affects execution; the remaining components assess recorded evidence offline.}
\label{tab:evaluators}
\centering\small
\begin{tabularx}{\linewidth}{@{}>{\raggedright\arraybackslash}p{24mm}>{\raggedright\arraybackslash}p{26mm}>{\raggedright\arraybackslash}X>{\raggedright\arraybackslash}X@{}}
\toprule
Component & Model & Inputs & Output / role\\
\midrule
System validator & Same as acting model & Criteria and execution evidence & Verdict and feedback controlling the loop\\
Validator-quality judge & Claude Sonnet 4.5 & Objective, evidence, explanation, reference-run context & Explanation-quality scores\\
Task planning judge & GPT-5 Codex & Candidate plans and local label definitions & Planning labels, severity, explanation\\
Trajectory panel & Five models listed in text & Experiment context and complete chat log & Progress, primary/contributing causes, mechanisms, recovery, severity\\
\bottomrule
\end{tabularx}
\end{table}

Every trajectory judge receives the experiment, environment, objective, available context, and complete cyber-agent log. It identifies the last completed stage and the point where meaningful progress becomes constrained, supporting its diagnosis with three to five evidence items tied to a subtask or timestamp. Instructions prohibit inferring credentials or privileges from a known username, automatically accepting the validator's conclusion, or mistaking superficial command variation for recovery.

Outputs follow a structured JSON (JavaScript Object Notation) schema containing progress, categories, tags, Boolean recovery and contribution assessments, severity from 1 to 5, and confidence from 0 to 1. Calls use temperature~0, at most 5,000 output tokens, and up to three attempts; JSON mode is requested where supported. Required fields and valid values are checked programmatically. Invalid, context-limited, or failed calls remain missing. Confidence is judge-reported metadata, not a vote weight or calibrated probability of correctness.

For trajectory $\tau$, let $n_c(\tau)$ be the number of valid judges selecting category $c$, and $J_\tau$ the number of valid judges. The modal category set and its support are
\begin{equation}
 \widehat{C}(\tau)=\operatorname*{arg\,max}_{c}n_c(\tau),\qquad
 A(\tau)=\frac{\max_c n_c(\tau)}{J_\tau}.
\end{equation}
Ties remain explicit; a unique mode can have only two of five votes and is not necessarily a majority. Setting-level mean support averages $A(\tau)$ over included trajectories. Mechanisms, adaptation, and contributions are descriptive judge-output counts nested within trajectories. They are not independent cyber-agent trials, so their denominators must not be used as an experimental sample size.

\section{Results}
\subsection{Validator Evidence Assessment}
Validator-explanation assessment uses expert-defined S1, covering 19 runs for successful explanations and 18 for unsuccessful attempts. These calibration runs include pilot configurations outside the principal benchmark. For each verdict, Claude Sonnet~4.5 receives the subtask objective, raw evidence, validator explanation, and corresponding evidence from a successful Claude Opus~4.5 reference run. The reference supplies semantic context, not human ground truth.

On a 0--1 scale, the overall explanation-quality score is 0.79 for successful attempts and 0.84 for unsuccessful attempts. Evidence grounding is 0.81 and 0.90, respectively, while specificity is lower at 0.69 and 0.68. Cautiousness differs more substantially: 0.70 for successful verdicts versus 0.92 for unsuccessful verdicts. These scores suggest that the external judge regarded success explanations as more prone to overstatement when evidence was incomplete. They assess explanations, however, and should not be interpreted as measured success-classification accuracy or proof that the validator was calibrated. No human-generated reference labels are available in this assessment.

\subsection{Localized Resource Amplification}
\begin{figure}[t]
\centering
\includegraphics[width=.4\linewidth]{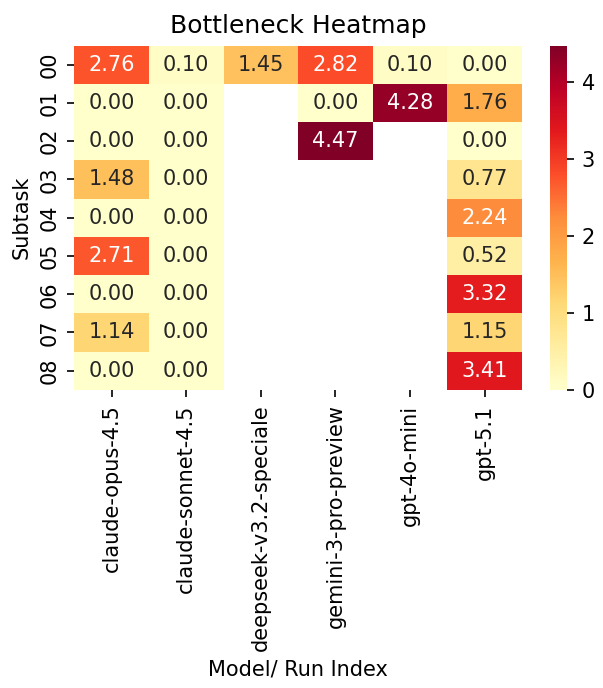}\hfill
\includegraphics[width=.33\linewidth]{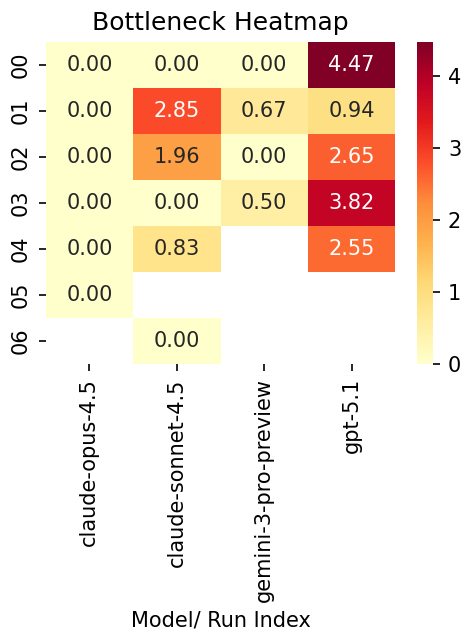}
\caption{Expert-defined task-level bottleneck scores, S1 (left) and S2 (right). Higher values indicate within-subtask resource amplification. Blank cells indicate subtasks the agent never reached, not zero scores. S2 shows seven of ten subtasks because no agent reached the final three. Blank cells indicate subtasks the agent never reached, not zero scores. S2 shows seven of ten subtasks because no agent reached the final three. GPT-4o-mini is absent from the S2 heatmap because it did not complete the first expert-defined subtask and therefore produced no successful-subtask record for task-level scoring, although its failed execution trace remained available for trajectory-level analysis.
}
\label{fig:heatmaps}
\end{figure}

Figure~\ref{fig:heatmaps} retains the expert-defined resource maps. S1 shows several isolated high-score cells, while S2 has repeated elevated scores in one plotted run and additional elevated cells elsewhere. These are within-subtask relative patterns, not evidence that one scenario is intrinsically harder or that every low-score execution is reliable. Missing cells must not be read as zero resource use.

The planning analysis associates difficult S1 cases with tool misalignment and fragile invocation structures; S2 also includes over-specification and missing environmental probing. Comparable resource scores receive different diagnoses, illustrating why localization and interpretation are complementary. We do not claim a statistically validated score--severity relationship.

\subsection{Primary Trajectory Diagnoses}
The revised result set contains 29 complete five-judge panels, or 145 judgments used in the aggregates. The Claude Sonnet~4.5 S1-A panel is now complete. Claude Opus~4.5 S2-A remains incomplete because a judge request was blocked by a provider-level cybersecurity safety filter. It is marked INC, rather than scheduled for rerun, and excluded from every reported aggregate. This is missing \emph{evaluation} data, not evidence that the acting model failed or that the trajectory belongs to a particular category.

Table~\ref{tab:coverage} reports 23 completed trajectories in the four-model comparison cohort. Under S1-E, action execution is the unique modal diagnosis for Opus, Gemini, and GPT-5.1, while Sonnet is tied between action execution and planning. Support varies considerably: Gemini's action-execution mode receives only 2/5 votes, whereas GPT-5.1 receives 5/5. The associated diagnoses concern command realization, quoting and escaping, credential-tool invocation, and lateral-agent deployment. Under S1-S, all four selected trajectories are planning-limited with 4/5 or 5/5 support. This is consistent with difficulty constructing and validating dependencies when decomposition is model-generated, but it is not a controlled estimate of scaffolding's causal effect.

S1-A is heterogeneous. Opus is recovery-limited after collecting but not operationalizing credential artifacts. Gemini is unanimously planning-limited because required credentials and a lateral payload are assumed rather than established. The completed Sonnet panel has a planning mode with 3/5 support. GPT-5.1 is tied between planning and framework limitations, with its trace ending before credential-inspection actions execute.

S2 reverses the simple ordering suggested by S1. Under expert decomposition, Opus and Sonnet have planning modes, Gemini a recovery mode, and GPT-5.1 a planning/recovery tie. The diagnoses emphasize unverified credentials, share permissions, and identity transitions. Under self-scaffolding, three trajectories have action-execution majorities; Opus is tied between execution and planning. The recurring blockage is agent deployment with a wrong executable, unresolved variables, placeholder commands, or acceptance of process creation as evidence of an operational callback. All three completed S2-A panels have planning modes with 3/5 support; the missing Opus panel prevents a complete four-model conclusion for this setting.

\subsection{Agreement, Adaptation, and Mechanisms}
Table~\ref{tab:summary} separates comparison-cohort agreement and adaptation from all-model severity and framework summaries. In S1, mean modal support is highest under self-scaffolding (0.90). Mean support is lower in S2, where prerequisite, command, and validation problems can coexist. Low support is substantive uncertainty, not merely a reporting inconvenience: some traces admit several plausible progress-limiting explanations.

A deterministic panel-deletion check is possible from the reported vote counts: 12 of the 23 completed comparison-cohort panels have at least 4/5 support, guaranteeing that their unique modal category survives removal of any one judge. This includes all four S1-S panels. The remaining 11 panels lack this guarantee from modal counts alone. This checks sensitivity to one deleted vote, not chance-corrected reliability, human correctness, or repeated-run robustness.

\begin{table}[t]
\caption{Setting summaries with explicit populations. Left block: completed panels in the four-model comparison cohort. Right block: all completed panels, including additional models. ``No adapt.'' counts judgments reporting no meaningful adaptation; ``contrib.'' counts validator/framework contributions. Severity uses the 1--5 scale. Neither block estimates repeated-run failure rates.}
\label{tab:summary}
\centering\small\setlength{\tabcolsep}{4pt}
\begin{tabular}{lcccccc}
\toprule
& \multicolumn{3}{c}{Comparison cohort} & \multicolumn{3}{c}{All models}\\
\cmidrule(lr){2-4}\cmidrule(lr){5-7}
Setting & Traces & Support & No adapt. & Traces & Severity & Contrib.\\
\midrule
S1-E & 4 & 0.65 & 5/20 & 6 & 4.03 & 22/30\\
S1-S & 4 & 0.90 & 15/20 & 6 & 4.83 & 29/30\\
S1-A & 4 & 0.70 & 20/20 & 5 & 4.68 & 18/25\\
S2-E & 4 & 0.65 & 20/20 & 5 & 4.92 & 20/25\\
S2-S & 4 & 0.65 & 20/20 & 4 & 5.00 & 20/20\\
S2-A & 3 & 0.60 & 12/15 & 3 & 4.80 & 15/15\\
\bottomrule
\end{tabular}
\end{table}

In the S1 comparison cohort, judgments reporting no meaningful adaptation increase from 5/20 under expert decomposition to 15/20 under self-scaffolding and 20/20 under full autonomy. The S1-S Opus trace is a useful counterexample: it abandons unsuccessful credential extraction, switches to password spraying, and completes the objective. Under S2-E and S2-S, all 20 judgments in each setting report no meaningful adaptation. Thus, a supplied task sequence does not by itself ensure that observed prerequisite failures lead to useful strategic revision. These are descriptions of the selected traces, not evidence that autonomy generally reduces recovery ability.

The mechanism counts in Table~\ref{tab:mechanisms} include all completed panels and are non-exclusive. S1-E emphasizes command construction (18/30), whereas S1-S emphasizes unsupported assumptions (25/30) and missing prerequisites (20/30). One recurring interpretation is that credentials associated with one domain identity are assumed to authenticate a different domain identity; changing execution tools then preserves the unresolved dependency. In S2-S, command-construction problems remain frequent even while unsupported assumptions persist. The broad categories therefore describe where judges place explanatory emphasis, not mutually isolated fault types.

Mean severity ranges from 4.03 to 5.00 across settings. Validator/framework contributions appear in 72--100\% of the included judgments, including 18/25 in S1-A. Examples include accepting file or process creation, or echoed output, as evidence of successful execution; advancing with unverified identities; terminating without an effective recovery path; and failing to retry malformed model outputs. Such judgments motivate checking the surrounding execution loop as well as the acting model. They do not quantify how much improvement a different validator would produce.

\section{Discussion and Conclusion}

\begin{table}[t]
\caption{Selected mechanism frequencies across all completed panels. CCE: command-construction error; UA: unsupported assumption; PRQ: missing or unverified prerequisite; SPD: single-point dependency; IR: insufficient reconnaissance. Counts are judge-output frequencies, not independent trajectory counts.}
\label{tab:mechanisms}
\centering\small
\begin{tabularx}{\linewidth}{@{}lX@{}}
\toprule
Setting & Most frequent reported mechanisms\\
\midrule
S1-E & CCE 18/30; UA 5/30; PRQ and SPD 3/30 each\\
S1-S & UA 25/30; PRQ 20/30; SPD and IR 11/30 each\\
S1-A & UA and PRQ 15/25 each; SPD 12/25; IR 8/25\\
S2-E & UA 18/25; PRQ 13/25; IR 12/25\\
S2-S & CCE 13/20; UA 12/20; PRQ 9/20\\
S2-A & UA and PRQ 13/15 each; SPD 6/15\\
\bottomrule
\end{tabularx}
\end{table}

The proposed framework combines subtask-level resource localization with trajectory-level diagnosis to identify where and why cyber-agent progress becomes constrained. Across 29 complete judge panels, the observed bottlenecks vary by scenario and scaffolding mode, spanning planning, execution, recovery, and framework effects rather than following a uniform trend with autonomy. The results motivate evaluation beyond final success, including verified progress, unresolved prerequisites, evidence interpretation, and whether retries reflect meaningful adaptation.

The findings should be interpreted as diagnostic evidence from selected high-performing trajectories rather than estimates of model-level failure prevalence. The study covers two lateral-movement scenarios in one environment, and LLM-judge agreement does not substitute for independent human validation. In addition, incomplete coverage and best-trajectory selection limit conclusions about within-model variability. Broader validation with repeated executions, additional scenarios, and expert annotations would strengthen the generality of the findings.

\section{Acknowledgement}
This project was conducted by the National Research Council of Canada, on behalf of the Canadian AI Safety Institute (CAISI).

\bibliographystyle{splncs04}
\bibliography{lib}

\end{document}